\documentclass[aps,prl,twocolumn,superscriptaddress,floatfix]{revtex4}

\usepackage{bm}
\usepackage{amsmath}
\usepackage{caption}
\usepackage{amssymb}
\usepackage{graphicx}
\usepackage{color}
\usepackage{soul,xcolor}
\setstcolor{red}
\usepackage{ulem}
\usepackage{cancel}

\newcommand\Ccancel[2][red]

\usepackage{xcolor}\definecolor{ao}{rgb}{0.0,0.0,1.0}

\definecolor{br}{rgb}{1.0, 0.22, 0.0}

\usepackage{epsfig}

\input{epsf}

\begin{document}

\title{Temporal evolution of fluxes in driven quantum dots}
\author{Debashree Chowdhury}
\affiliation{Centre of Nanotechnology, Indian Institute of Technology Roorkee, Roorkee, Uttarakhand-247667}
\email{debashreephys@gmail.com}
\date{\today}
\begin{abstract}
Driven mesoscopic system is a topic of great recent interest. The temporal evolution of the fluxes(particle and energy) are studied in a system of a driven single level quantum dot. At a very low reservoir temperature $T\rightarrow 0$ and for common chemical potentials of the two reservoirs, we have presented analytical expressions for time dependent particle and energy fluxes in a very simple form. Apart from these fluxes, the behavior of the dot occupation and the power developed in the system due to the presence of the time dependent drive are also being studied. Importantly, for a very low frequency of the drive, one finds a directed energy flow towards the leads. Increasing the frequency from low to medium, one finds change in the direction of the energy flow depending on the time. These results can also be verified experimentally.
\end{abstract}

\maketitle
\section{Introduction}
The most significant tool for the theoretical framework of the electron transport in mesoscopic domain is the non equilibrium Green's function formalism\cite{1,2}. Few previous works \cite{3,4,5,6} have boosted up interesting applications of this Green's function method in the mesoscopic transport issues. In recent times the quantum transport in presence of time periodic fields has gained considerable attention. Different experimental studies in presence of periodic driving fields and experiments on pumping phenomena \cite{7,8,9,10,11,12,13} have encouraged the theoreticians to formulate and solve the so called Dyson's equations in theoretical models for different driven mesoscopic systems. Time periodic fields are useful in controlling matter tunneling in Bose-Einstein condensates\cite{16a}. In Refs. \cite{19a,20a}, the authors have analyzed the production of heat in nanoscale engines. Similar analysis for molecular heat pumping is proposed in Ref. \cite{21a}. Importantly, in presence of the time periodic fields, directed flow of charge, spin and energy can be observed in the system \cite{14a,15a}. In Ref.\cite{14}, the electron transport is analyzed for a system of mesoscopic ring threaded by a time-dependent magnetic flux. On the other hand, analysis is also being done in models of quantum pumps having a quantum dot in presence of ac gate voltages \cite{15,16,17}.

In our recent papers, \cite{22,23} we have dealt with a mesoscopic system driven by two different time dependent fields: random stochastic and the time periodic electric field. In these papers we have computed time/noise averaged particle and energy fluxes for a very simple set up. The telegraph noise is the simplest model for the environment \cite{22,23}, where the noisy environment is modeled by a single nearby impurity, which jumps randomly between two states. This eventually induces a time dependent random potential on the system. In this scenario, the average particle and energy fluxes are analyzed\cite{22}.
 \begin{figure}\label{fig1}
	\includegraphics[width=0.8 \linewidth]{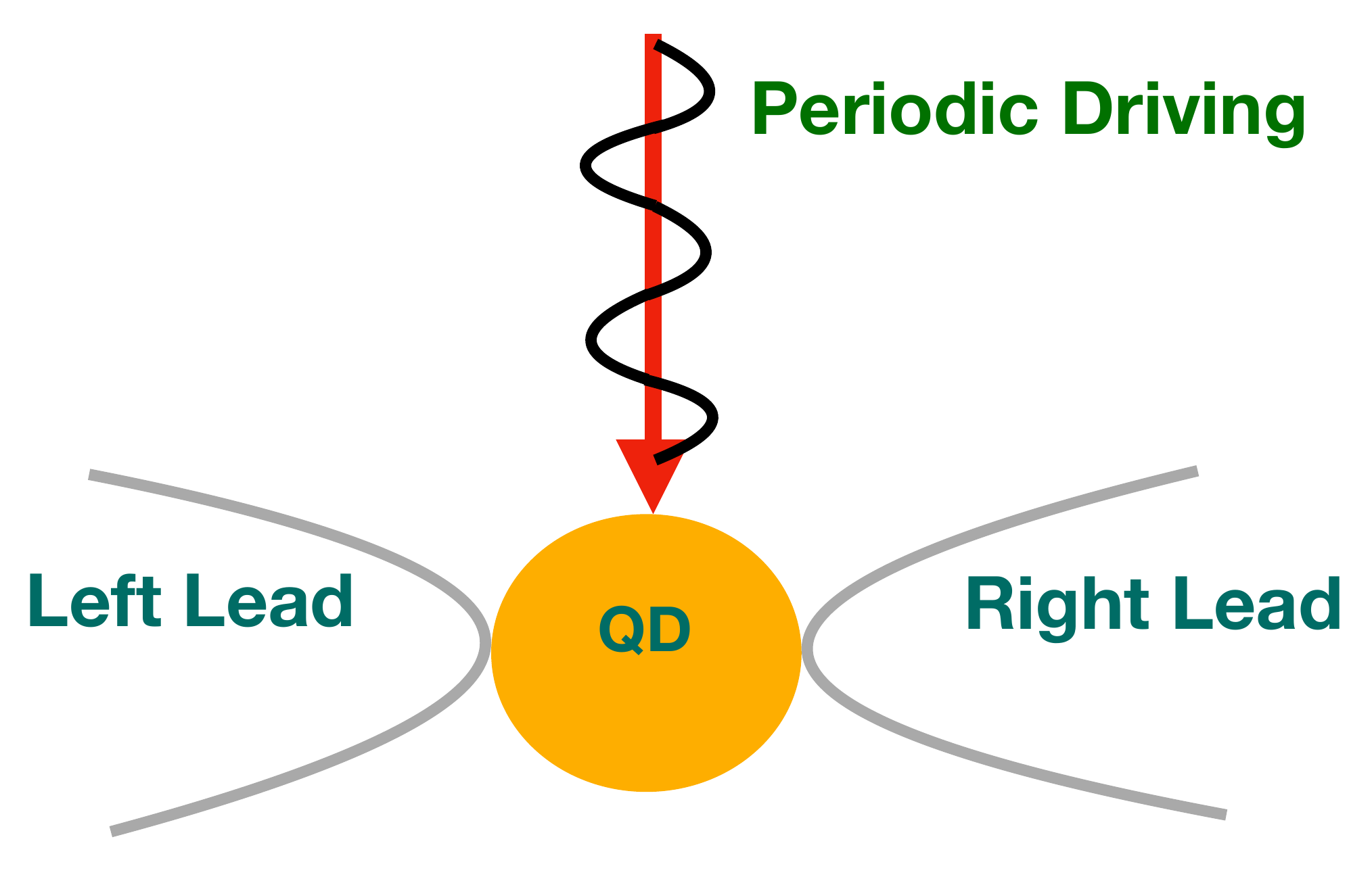}
	\caption{Schematic diagram of the set up. }
\end{figure}
 The end results for both the cases show modification in the conductivity due to the presence of time dependent fields. It is to be mentioned here that the results for these two cases are more or less alike, however the difference lies in the averaging technique. While noise averaging is considered for the telegraph noise case i.e averaging over all histories of time \cite{22}, we use averaging over a single period for the periodic case\cite{23}. In our recent work \cite{23}, the main focus was on the time averaged quantities over one time period for the periodic driving. As a consequence we end up with results having no time dependence at all. 
 The motivation of the present paper is to examine the explicit time dependence in various quantities and the time evolution of different fluxes (particle and energy). The system used in this paper is a single level quantum dot(QD), which is connected to two fermionic reservoirs via two metallic leads. The dot is further affected by an external periodic drive. The schematic diagram for the system is presented in FIG. 1. The goal of this paper is also to see the time evolution of the important quantities such as the dot occupation and power developed by the source of time dependence when the two reservoirs are kept at the common temperature and common chemical potentials. The novelty of the current paper is to have results for the fluxes(both particle and energy) in a simple form. The fluxes usually contain complicated integrals over energy. As the results of this paper are evaluated in the limit of $T\rightarrow 0,$ and with common chemical potentials of the two reservoirs, the integrals boil down to a simple form. As a consequence, one gets simplified expression of fluxes, which could be analyzed for different frequency values of the drive. It is shown here that the direction of the lead energy fluxes in the system for a low frequency drive is always from the source of the time dependence towards the leads. In the moderate frequency of the drive, the direction of energy flow becomes fluctuating depending on the value of $t.$

In this paper, we are interested in the behavior of different fluxes in presence of time periodic fields, which renormalizes the energy of the dot as follows
\begin{align}\label{1}
\epsilon_{d}(t)=\epsilon+ U \cos(\Omega t),
\end{align} 
where $\epsilon$ is the bare energy of the electrons within the dot, $\Omega$ is the frequency of the periodic electric field. It is considered here that the periodic electric field is applied only on the central part of the system (quantum dot) and no other parts are affected.
As a consequence of such a modification on the energy of the dot, the Hamiltonian of the system becomes time dependent as follows
\begin{align}\label{2}
&H(t)= \epsilon^{}_{d}(t)d^{\dagger}_{}d+\sum_{\bf k}(V^{}_{\bf k}c^{\dagger}_{\bf k}d+{\rm Hc})+\sum_{\bf p}(V^{}_{\bf p}c^{\dagger}_{\bf p}d+{\rm Hc})\nonumber\\&+\sum_{\bf k}\epsilon^{}_{k}c^{\dagger}_{\bf k}c^{}_{\bf k}+
\sum_{\bf p}\epsilon^{}_{p}c^{\dagger}_{\bf p}c^{}_{\bf p}.
\end{align}  
The first term on the right hand side of Eq. (\ref{2}) is the system Hamiltonian, the second and third terms are the tunneling terms and the last two terms are the Hamiltonian of the leads(left and right leads). Here in Eq. (\ref{2}), $d$ ($d^{\dagger}$) is the annihilation(creation) operator of electrons on the level. The dot is coupled to two electronic reservoirs of spin-less electrons by tunneling amplitudes $V_{\bf k}$ and $V_{\bf p}.$ $c^{}_{{\bf k}({\bf p})}$ and $c^{\dagger}_{{\bf k}({\bf p})}$ are the annihilation (creation) operators for the electrons in the two leads and Hc denotes the hermitian conjugate. We use the wave vector ${\bf k}$ (${\bf p}$) for the states on the left (right) lead. Using Eq. (\ref{2}), we would discuss the generation of different currents in the system in the next section.

We organize our analysis as follows: In Sec. II we have reviewed the time dependent fluxes in terms of the Keldysh Green's functions for a general time dependent Hamiltonian. The detailed discussion of the fluxes in presence of the periodic electric field is presented in Sec III. Sec IV contains the main results of this paper, which includes fluxes for $T\rightarrow 0$ with a common chemical potential of the reservoirs and we conclude in Sec. V.  
\section{Different fluxes}
 A consistent way to define the particle current of the lead is to take the time derivative of the electronic density of that particular lead. The time derivative of different parts of $H(t)$ provide different energy currents corresponding to that term\cite{ref1,ref2}. The goal here is to analyze these particle and energy fluxes in the time domain.

From Eq. (\ref{7}), the particle current flux can be obtained in the following form \cite{ref1,ref2}
\begin{align}\label{19}
I^{}_{L}(t)&=\langle \frac{d}{dt}\sum_{\bf k}c^{\dagger}_{\bf k}c^{}_{\bf k}\rangle\nonumber\\& =2\Gamma_{L}Q_{d}(t)+4\int \frac{d\omega}{2\pi}\Gamma_{L}f_{L}(\omega){\rm Im}[G_{dd}^{r}(t,\omega)],
\end{align}

where we have denoted the occupation of the dot as $Q_{d}(t), $ which is defined in Eq. (A14). Importantly, we consider $\hbar=1$ in all calculations of the paper. Here
\begin{equation}\label{Gamma}
\Gamma=\Gamma^{}_L+\Gamma^{}_R\ ,\ \ \ \Gamma^{}_{L(R)}=\pi {\cal S}^{}_{L(R)}|V^{}_{L(R)}|^2
\end{equation}
(${\cal S}^{}_{L(R)}$ is the density of states in left(right) reservoir. Importantly, it is assumed that the tunneling amplitudes are determined by the wave vectors corresponding to the energy of the tunneling electrons and thus the tunneling amplitudes are considered to be independent of the wave vectors, which is coined as the wide-band approximation \cite{22,23,ref1,ref2}. $f_{L}(\omega)=\frac{1}{1+e^{\beta(\omega-\mu_{L})}}$ is the Fermi distribution function of the left lead electrons with $\mu_{L}$ as the chemical potential of the left reservoir and $\beta=1/KT,$ where $K$ is the Boltzmann constant and $T$ is the temperature of the left reservoir .
Substituting $Q^{}_{d}(t)$ from Eq. (\ref{18}) to Eq. (\ref{19}), we get (see appendix A for detailed calculations)
\begin{align}\label{20}
&I^{}_{L}(t)=-\frac{\Gamma_{L}}{\Gamma}\frac{dQ_{d}(t)}{dt}\nonumber\\&+\frac{4\Gamma_{L}\Gamma_{R}}{\Gamma}\int \frac{d\omega}{2\pi}{\rm Im}[G_{dd}^{a}(t,\omega)]\Big(f_{R}(\omega)-f_{L}(\omega)\Big).
\end{align}
One can easily check that the flux of particles into the dot is compensated by the sum of the two fluxes into the leads i.e,
\begin{align}\label{9}
I^{}_{d}(t)=\frac{d}{dt}Q_{d}(t)=-[I^{}_{L}(t)+I^{}_{R}(t)]\ ,
\end{align}
which is nothing but the particle current conservation.

Apart from charge flux we are also interested in different energy fluxes that flow in the time domain\cite{ref1,ref2}.
From the time derivative of the Hamiltonian of the left lead, one gets the energy flux into the left reservoir as \cite{ref1,ref2}
\begin{align}
&I^{E}_{L}(t)
=\Big\langle \frac{d}{dt}\sum_{\bf k}\epsilon^{}_{k}c^{\dagger}_{\bf k}c^{}_{\bf k}\Big\rangle\nonumber\\
&=4\Gamma^{}_{L}
\int\frac{d\omega}{2\pi}\Sigma(\omega){\rm Re}[ G_{dd}^{r}(t,\omega)]\nonumber\\&-8\Gamma^{}_{L}
\int\frac{d\omega}{2\pi}\Sigma(\omega)
\epsilon_{d}(t)\int^{t}_{}dt' e^{2\Gamma (t'-t)} {\rm Im}[ G_{dd}^{r}(t',\omega)] \nonumber\\&+4\Gamma_{L}\int\frac{d\omega}{2\pi} f_{L}(\omega)\omega{\rm Im}[ G_{dd}^{r}(t,\omega)],
\label{ILE}
\end{align}
where $\Sigma(\omega)=[\Gamma^{}_{L}f^{}_{L}(\omega)+\Gamma^{}_{R}f^{}_{R}(\omega)].$ 
The energy flux associated with the right lead is derived from Eq. (\ref{ILE})
by interchanging $L\Leftrightarrow R$ and ${\bf k}\Leftrightarrow {\bf p}$.
Another part of the energy flux can be obtained as
\begin{align}\label{11}
I_{d}^{E}(t)&=\frac{d}{dt}[\epsilon_{d}(t)Q_{d}(t)]\nonumber\\
&=\frac{d\epsilon_{d}(t)}{dt}Q_{d}(t)+\epsilon_{d}(t)I_{d}(t).
\end{align}
In Eq. (\ref{11}), the first term on the right hand side is the power supplied by the source of time dependence to the system. The power is denoted as
\begin{align}\label{12}
P_{d}(t)=Q_{d}(t)\frac{d\epsilon_{d}(t)}{dt}
\end{align}
in the rest of the paper.
Also from Eqs. (\ref{ILE}),(\ref{11}) and (\ref{12}), we have \cite{ref1,ref2}
\begin{align}
I_{L}^{E}(t)+I_{R}^{E}(t)+I_{tun,L}^{E}(t)+I_{tun,R}^{E}(t)+I_{d}^{E}(t) = P_{d}(t),
\end{align}
which expresses the energy conservation through the junction.

The temporal variation of the (left and right) tunneling Hamiltonian gives \cite{ref1,ref2,22,23}
\begin{eqnarray}
&I^{E}_{{\rm tun}, L}(t)= \langle \frac{d}{dt}\sum_{\bf k}(V^{}_{\bf k}c^{\dagger}_{\bf k}d+{\rm Hc})\rangle\nonumber\\
&=4\Gamma_{L}\int\frac{d\omega}{2\pi} f^{}_{L}(\omega)(-\omega+\epsilon_{d}(t)) {\rm Im}[ G_{dd}^{r}(t,\omega)]\nonumber\\&-4\Gamma^{}_{L}\Gamma
\int\frac{d\omega}{2\pi} f_{L}(\omega){\rm Re}[ G_{dd}^{r}(t,\omega)] .
\label{ITLE}
\end{eqnarray}

In Refs\cite{ref1,ref2,22}, it is shown that the time-averaged tunneling energy flux  vanishes, but this energy flux can be visualized as a temporary energy storage\cite{ref1,ref2,lil,lil1}. Thus the tunneling region can be called as an energy reactance, which modulates peak power developed in the dynamics \cite{lil}.

\section{Analysis of fluxes in presence of periodic driving }
In this section, we are interested in the results of a periodic driving on our simple system. As it is evident that all the fluxes can be derived when we have the complete knowledge of the Green's functions of the dot.
The retarded (advanced) Green's function on the dot is
\begin{equation}
G^{r(a)}_{dd}(t,t')=\mp i \Theta (\pm t\mp t')
e^{-i\int_{t'}^{t}dt^{}_{1}\epsilon^{}_{d}(t^{}_{1})\mp \Gamma (t-t')}
\ .
\label{GDRA}
\end{equation}
As is evident from (\ref{GDRA}), both the retarded and advanced dot Green's functions depend on the time-dependent energies $\epsilon^{}_d(t)$ only through the function
$
{\mathfrak X}(t,t')=e^{i\int_{t'}^t dt''\epsilon_d(t'')}\ .
$
In presence of a periodic drive, the energy of the dot is renormalized as given in Eq. (\ref{1}). Setting $\epsilon$ as the zero of energies, we have 
\begin{align}
{\mathfrak X}(t,t')&=\sum_{n,m=-\infty}^\infty J^{}_n(x)J^{}_m(x)e^{i\Omega(n t-mt')}\ ,
\label{Xper}
\end{align}
where $J^{}_n(x)$ is the Bessel function of the first kind and $x\equiv U/\Omega,$ where $U$ and $\Omega$ are the strength and frequency of the oscillatory field.
It follows from Eq. (\ref{GDRA}) and (\ref{Xper})
\begin{align}\label{26} 
&G^{r}_{dd}(t,\omega)
&=\sum_{m=-\infty}^{\infty}\sum_{n=-\infty}^{\infty}\frac{J^{}_{n}(x )J^{}_{m}(x )}{\alpha^{}_{-n}}e^{i(n-m)\Omega t^{}_{}}\ ,
\end{align}

where
\begin{align}\label{28}
\alpha^{}_{n}&=\omega+n\Omega +i\Gamma\ ,
\end{align}
and $G^{a}_{dd}(t,\omega)=[G^{r}_{dd}(t,\omega)]^{*}$ is the advanced Green's function of the dot. 
From ref. \cite{4}, one finds that it is useful to define the spectral function, which is related to the advanced Green's function as follows
\begin{align}\label{A}
&{\cal A} (t,\omega)=2{\rm Im}[G_{dd}^{a}(t,\omega)].
\end{align}
The spectral function can be interpreted as the available density
of states which are filled up according to the Fermi function to obtain the electron density.
The spectral function appears in Eqs. (\ref{19}), (\ref{ILE}) and (\ref{ITLE}) through the definition (\ref{A}). Thus it is worth mentioning that the analysis of this quantity is an crucial aspect of this paper. One can analyze the behavior of the spectral function when different parameters of the system are varied. It is important to see that in FIG.2 the variation of ${\cal A}(t,\omega)$  with $\omega$ is rapid for small $\omega$ values but very slow for larger values of $\omega.$ Another observation is that with increasing $U$ the amplitudes of the peaks become smaller and the peak shifts towards the right for $t=5,$ whereas it shifts towards left for $t=10$. It is important to note that we have chosen the values (t=5,10) to see the effect in the time interval of five. One can take any other periodicity(say for example t=1 or 2 ), the results will be the same as explained above. Thus a periodicity can be observed when time is varied. The reason behind the diminishing amplitude is that as we change $U,$ the values of the Bessel's functions also get modified and we have different amplitudes for different $U$s. Another important observation here is that in FIG. 2, we have considered that the frequency of the drive is much larger than the parameter $\Gamma.$ As we go to the lower frequency range, the plots will be unaltered as one increases time. We will further explain this point in the next section.

Let us now move towards the analysis of the dot occupation and the power supplied to the system. The occupation of the dot is\cite{22,23}
	\begin{align}\label{33}
&Q_{d}(t)=2\int\frac{d\omega}{2\pi}\Big[\Gamma_{L}f_{L}(\omega)+\Gamma_{R}f_{R}(\omega)\Big]\Big[{\mathfrak K}(t,\omega)+cc\Big],
\end{align}
where 
\begin{equation}
{\mathfrak K}(t,\omega)\equiv \int_{-\infty}^t dt^{}_1 e^{2\Gamma(t^{}_1-t)}\int_0^\infty d\tau e^{-(\Gamma+i\omega)\tau}{\mathfrak X}(t^{}_1,t^{}_1-\tau)\ 
\label{KKK}
\end{equation}


From Eq. (\ref{12}), the power supplied to the system from the source of the time periodic electric field can be obtained as\cite{22,23}
\begin{align}\label{37}
P_{d}(t)=2\int \frac{d\omega}{2\pi}&\Big[\Gamma_{L}f_{L}(\omega)+\Gamma_{R}f_{R}(\omega)\Big]\nonumber\\&\times\Bigg[\frac{d\epsilon_{d}(t)}{dt}{\mathfrak K}(t,\omega)+cc\Bigg].
\end{align}
To examine the temporal variation of $P_{d}(t)$ we need the analysis of ${\rm Re}\Bigg[\frac{d\epsilon_{d}(t)}{dt}{\mathfrak K}(t,\omega)\Bigg].$
Using Eq. (\ref{Xper}) in (\ref{KKK}) we proceed to find simplified expression for (\ref{33}) and (\ref{37}).

\begin{figure}\label{fig3}
	\includegraphics[width=0.9 \linewidth]{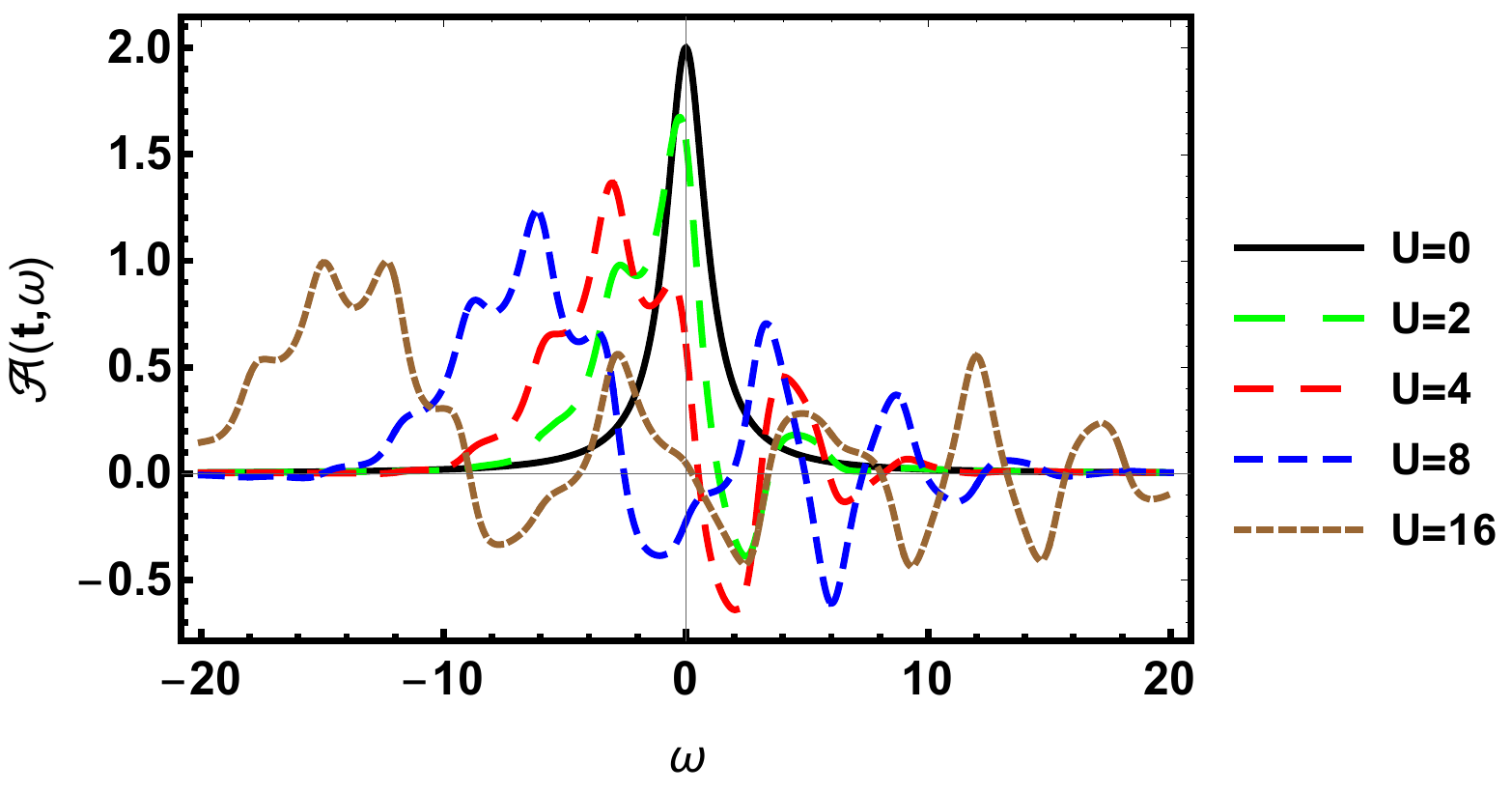}
	\includegraphics[width=0.9 \linewidth]{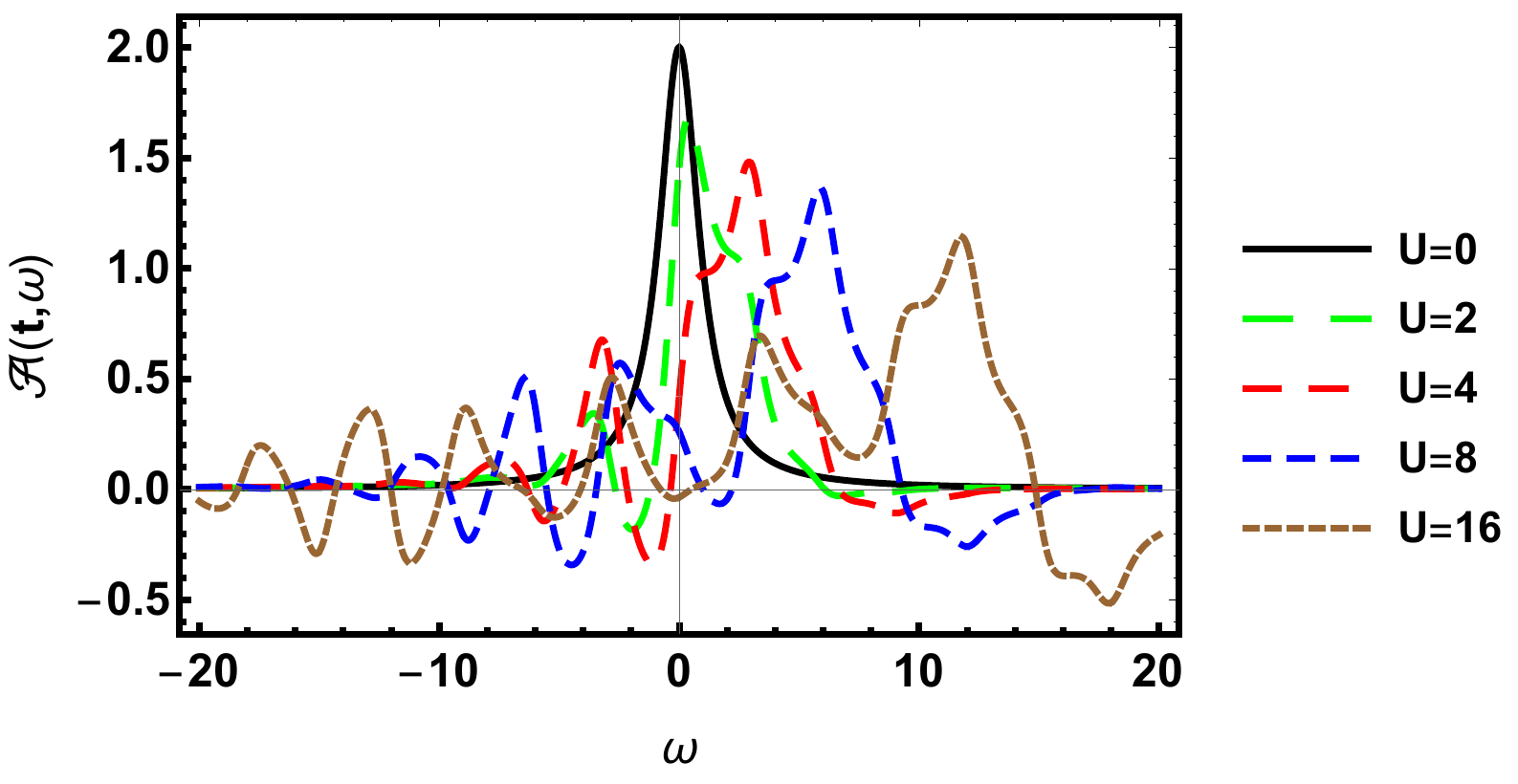}
	\caption{Plot of ${\cal A} (t,\omega)$ vs $\omega,$ for two different values of time ($t=5({\rm up}),10({\rm down})$). The black solid line in each plot is for $U=0$ and other lines with decreasing dashes are for U=2(Green),4(Red),8(Blue),16(Brown) with $\Omega=3.$ All quantities are in the unit of $\Gamma.$ The unit of $\Omega$ is same to the unit of $\Gamma$ and the unit of $t$ is same to that of $2\pi/\Omega.$}	
\end{figure}

\begin{figure}\label{fig4}
	\includegraphics[width=0.9 \linewidth]{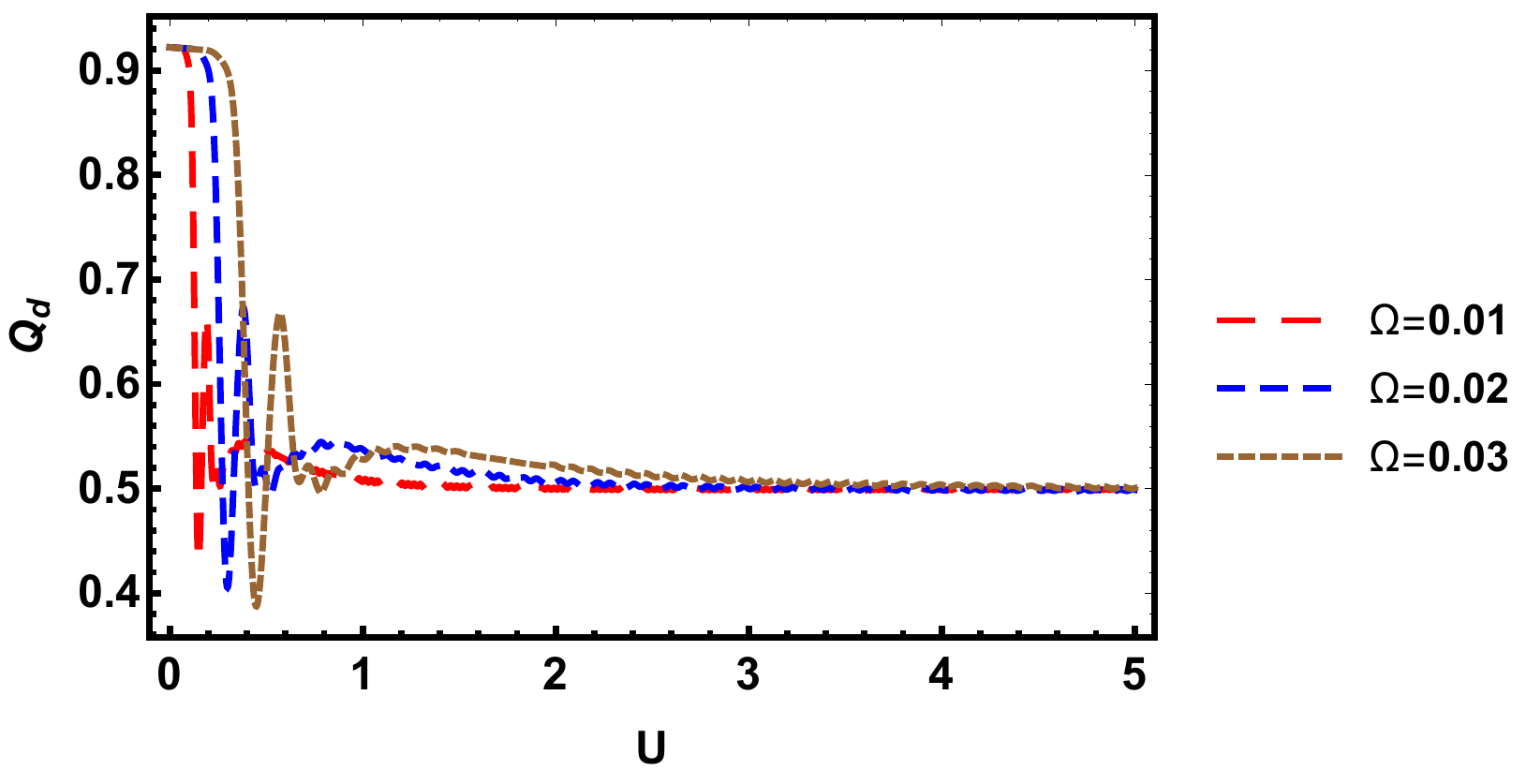}
	\hspace{.01 cm}
	\includegraphics[width=0.9 \linewidth]{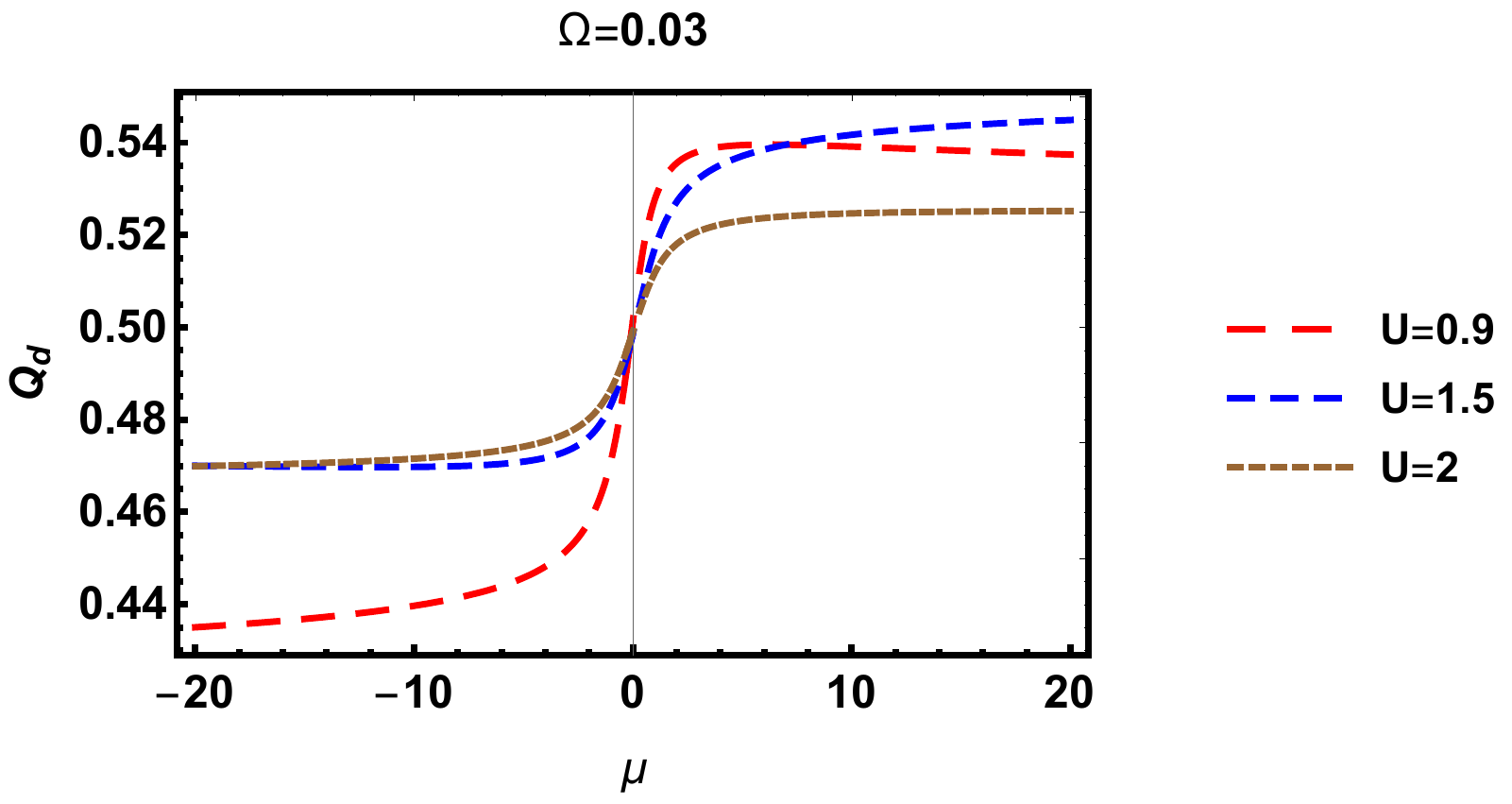}
	\caption{Up: Plot of the occupation of the dot with $U$ for three different $\Omega=0.01,0.02,0.03$. Down: Plot of the occupation of the dot with common chemical potential of the reservoirs $\mu$ and $\Omega =0.03$ for various $U=0.9,1.5,2.$  All quantities are in the unit of $\Gamma.$  The unit of $\Omega$ is same to the unit of $\Gamma$ and the unit of $t$ is same to that of $2\pi/\Omega.$}
\end{figure}

\section{Results for $T \longrightarrow 0$}

This section contains the main results of the paper. Here we are interested to discuss more on the fluxes, occupation number and power developed to the system when we are in a very low temperature domain. From Eq. (\ref{33}) using Eqs.  (\ref{Xper}) and (\ref{KKK}) we write the occupation number of the dot for the common chemical potential and common temperature of the two reservoirs(i.e $f_{R}(\omega)=f_{L}(\omega)=f_{}(\omega)$) as 
\begin{align}\label{Qd}
&Q_{d}(t)=4\Gamma\sum_{n}\sum_{m}J_{m}(x)J_{n}(x)\int^{t}_{-\infty} dt_{1}\nonumber\\& e^{2\Gamma(t_{1}-t)}\int_{-\infty}^{\infty}\frac{d\omega}{2\pi} f(\omega)\Big[\frac{\Gamma \cos[\Omega(n-m)t_{1}]}{\Gamma^{2}+(\omega-m\Omega)^{2}}\nonumber\\&+\frac{(\omega-m\Omega)\sin[\Omega(n-m)t_{1}]}{\Gamma^{2}+(\omega-m\Omega)^{2}}\Big],
\end{align} 
where $f(\omega)$ is the common Fermi distribution function of the reservoirs.
One can use the following
\begin{align}
\int_{-\infty}^{\infty} d\omega f(\omega)F^{'}(\omega)=-F(-\infty)+\int_{-\infty}^{\infty} d\omega[-f^{'}(\omega)] F(\omega),
\end{align}
in order to solve Eq. (\ref{Qd}). Here prime($'$) denotes the derivative with respect to $\omega$.  At low enough reservoir temperature, $[-f^{'}(\omega) ]=\delta(\omega).$ Although the value of $\log[\Gamma^{2}+(\omega-m\Omega)^{2}]$ diverges at $\omega\rightarrow -\infty,$ using
\begin{align}
\sum_{m}\sum_{n}J_{m}(x)J_{n}(x) e^{i\Omega (n-m)t}=1,
\end{align}
we can simplify the expression of the dot occupation. Thus for $T\rightarrow 0$ and for common chemical potential  $\mu_{L}=\mu_{R}=\mu$ of the reservoirs, the dot occupation can be written in a compact form as

\begin{figure}\label{fig5}	
	\includegraphics[width=0.9 \linewidth]{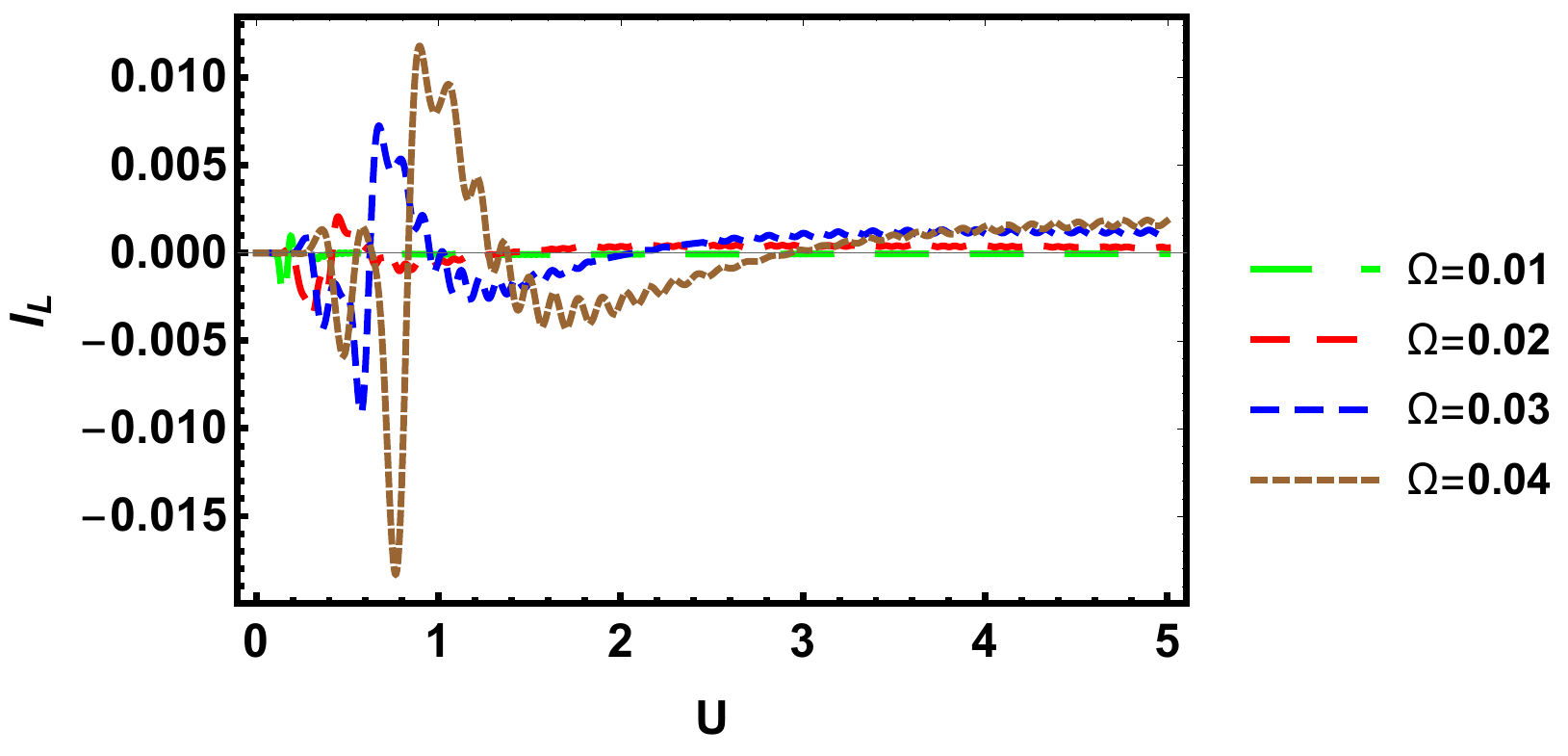}
	\caption{Up: Plot of the left lead particle flux with U for different $\Omega$s.  The unit of $\Omega$ is same to the unit of $\Gamma$ and the unit of $t$ is same to that of $2\pi/\Omega.$}	
\end{figure}
\begin{align}\label{43}
Q_{d}(t)&=\alpha^{Q}_{0}+\sum_{l}\Big(\alpha^{Q}_{l}\cos[l\Omega t]-\beta^{Q}_{l}\sin[l\Omega t]\Big),
\end{align}
where $l=(n-m).$ For different values of $l, $ we have different harmonics corresponding to $\cos(\Omega t),\cos(2\Omega t),\cos(3\Omega t)$ and also $\sin(\Omega t),\sin(2\Omega t),\sin(3\Omega t)$ etc. In (\ref{43}) we have redefined the parameters as follows
\begin{align}\label{alphaQ}
\alpha^{Q}_{0}&=1/2+\frac{2\Gamma}{\pi}\sum_{m}\alpha_{mm}\nonumber\\
\alpha^{Q}_{l}&=\frac{2\Gamma}{\pi}\sum_{m}\alpha_{m+l,m}\nonumber\\
\beta^{Q}_{l}&=\frac{2\Gamma}{\pi}\sum_{m}\beta_{m+l,m};
\end{align}
with
\begin{align}\label{42}
&\alpha_{nm}=\frac{J_{m}(x)J_{n}(x)}{(4\Gamma^{2}+(m-n)^{2}\Omega^{2})}\times\nonumber\\&\Big[\frac{(m-n)\Omega}{2}\log(\Gamma^{2}+(\mu-m\Omega)^{2})+2\Gamma\arctan(\frac{\mu-m\Omega}{\Gamma})\Big]\nonumber\\
&\beta_{nm}=\frac{J_{m}(x)J_{n}(x)}{(4\Gamma^{2}+(m-n)^{2}\Omega^{2})}\times\nonumber\\&\Big[-\log(\Gamma^{2}+(\mu-m\Omega)^{2})+(m-n)\Omega\arctan(\frac{\mu-m\Omega}{\Gamma})\Big].
\end{align}
 On the right side of Eq. (\ref{43}), the term $\alpha_{0}^{Q}$ is nothing but the averaged value of the occupation number for a single time period as is obtained in \cite{23}(see Eq. (73)\cite{23}) and is given as
\begin{align}
\alpha^{Q}_{0}=1/2+\frac{2\Gamma}{\pi}\sum_{m}\frac{J_{m}^{2}(x)}{2\Gamma}\arctan[\frac{\mu-m\Omega}{\Gamma}].
\end{align}
In FIG.3, the behavior of the occupation of the dot is plotted with $U$(Up) and $\mu$(Down). The up plot shows the behavior of the dot occupation for very small frequency of the external field. Note that $Q_{d}$ varies very sharp for small $U$ values but saturates to the time averaged value(0.5) when $U$ is increased. This plot remains unaltered as we increase time. The down plot shows plot of the occupation of the dot with $U$ for moderate frequency($\Omega=0.3$). Here one observes that plots for different $U$s meet at $\mu=0$ and the corresponding value is $Q_{d} =0.5$(the average value of $Q_{d}.$) This behavior changes as we go to the high frequency regime of the drive and /or change time.  
\begin{figure}\label{fig5a}	
	\includegraphics[width=0.9 \linewidth]{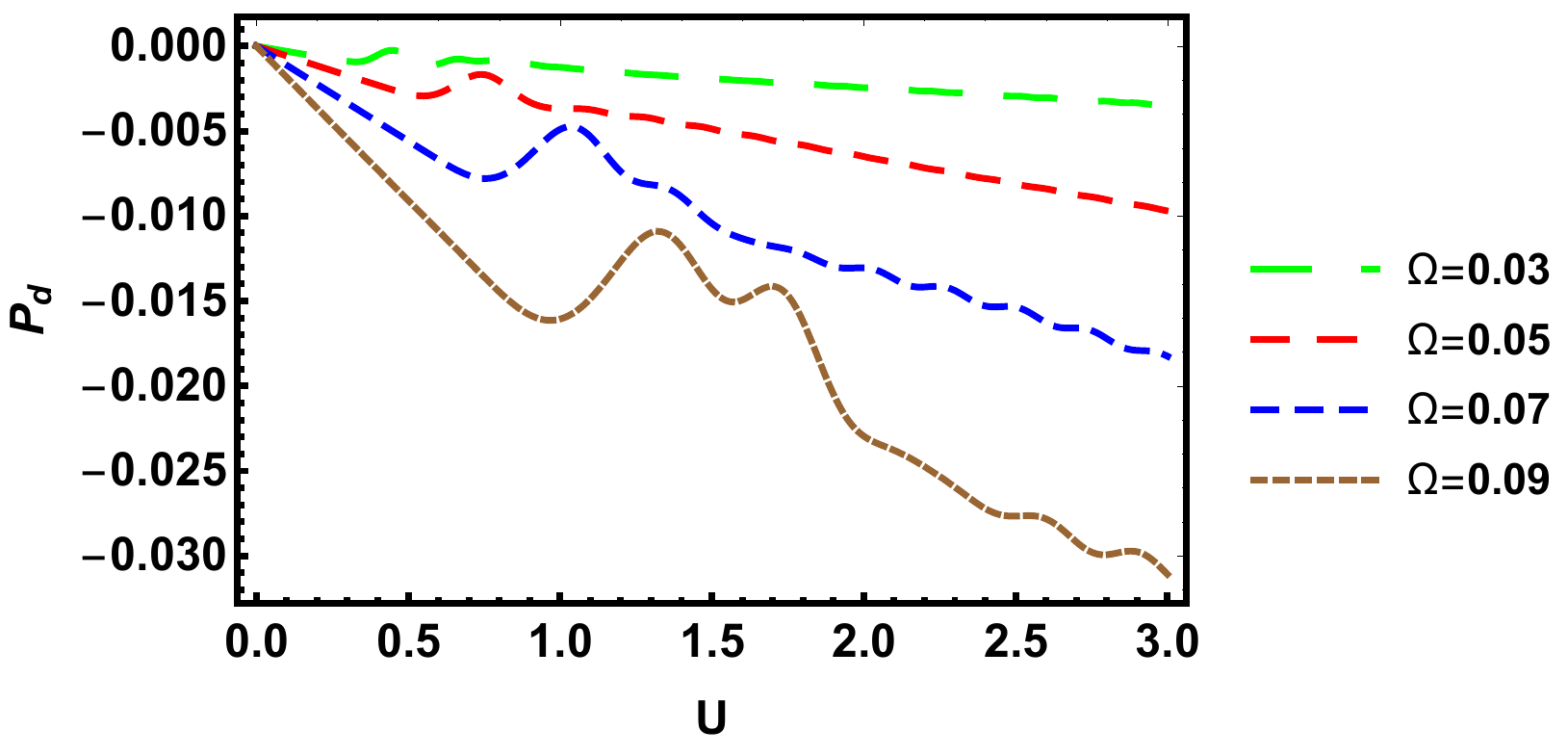}
	\hspace{.01cm}
	\includegraphics[width=0.9 \linewidth]{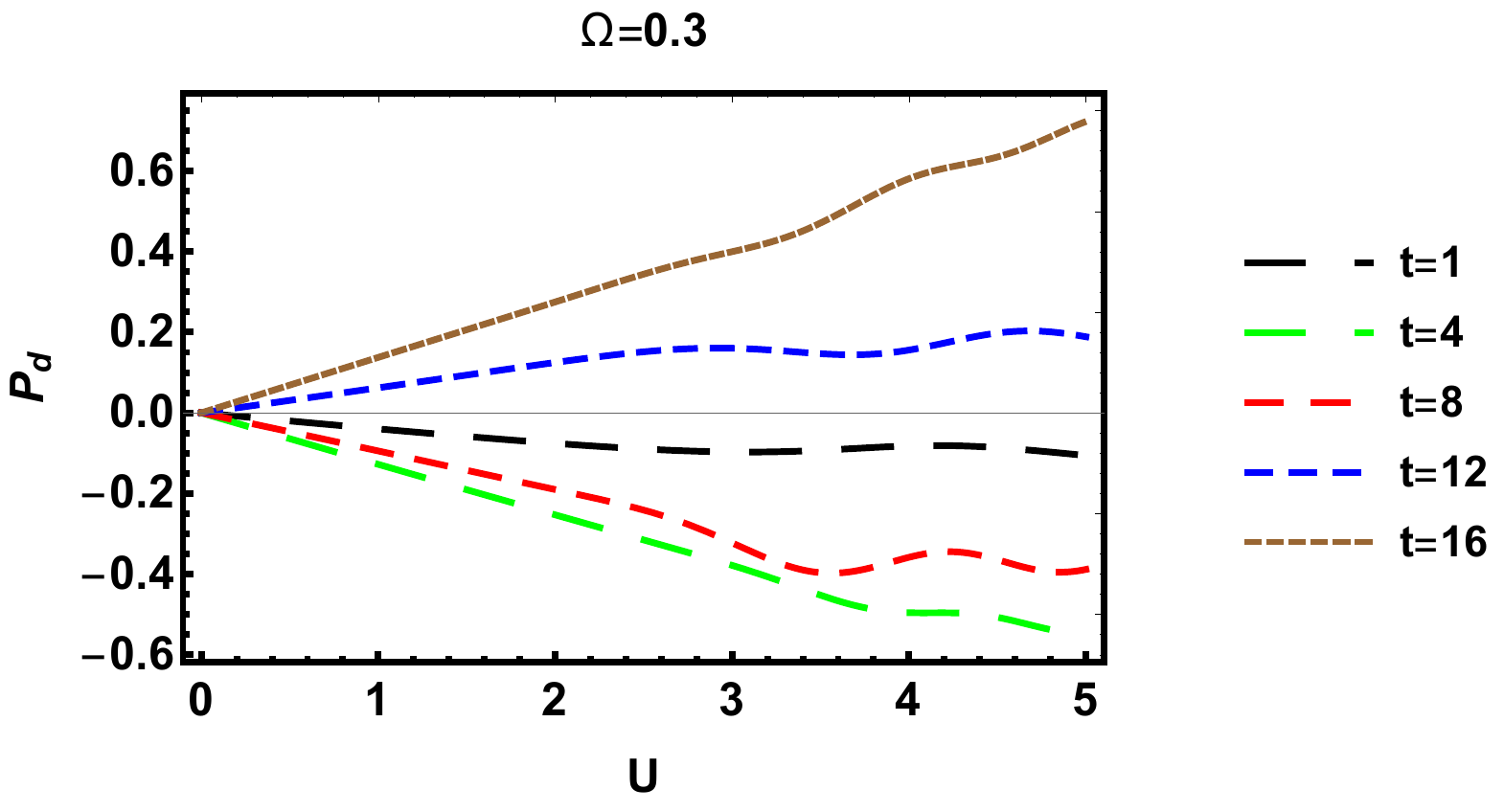}
	\caption{Up: Plot of the power delivered by the source of the periodic field with $U$ for $\mu =4,$ $t=5$ and for very small frequencies(0.01,0.02,0.03,0.04,0.05). All quantities are in the unit of $\Gamma.$ Down: Plot of the power with $U$ for moderate frequency $\Omega=0.3$ and for different times $t=1,4,8,12,16$.  The unit of $\Omega$ is same to the unit of $\Gamma$ and the unit of $t$ is same to that of $2\pi/\Omega.$}	
\end{figure} 

The particle flux of the left lead for $f_{L}(\omega)=f_{R}(\omega)=f_{}(\omega)$ is (see Eq. (\ref{20}))
\begin{align}\label{52}
&I_{L}(t)=-\frac{\Gamma_{L}}{\Gamma}\frac{dQ_{d}(t)}{dt}\nonumber\\
&=\frac{\Gamma_{L}\Omega}{\Gamma}\Big[\sum_{l}l\Big(\alpha^{Q}_{l}\sin[l\Omega t]+\beta_{l}^{Q}\cos[l\Omega t]\Big)\Big]
\end{align}
Eq. (\ref{52}) gives the expression of time dependent particle flux.
\begin{figure}\label{fig6}	
	\includegraphics[width=0.9 \linewidth]{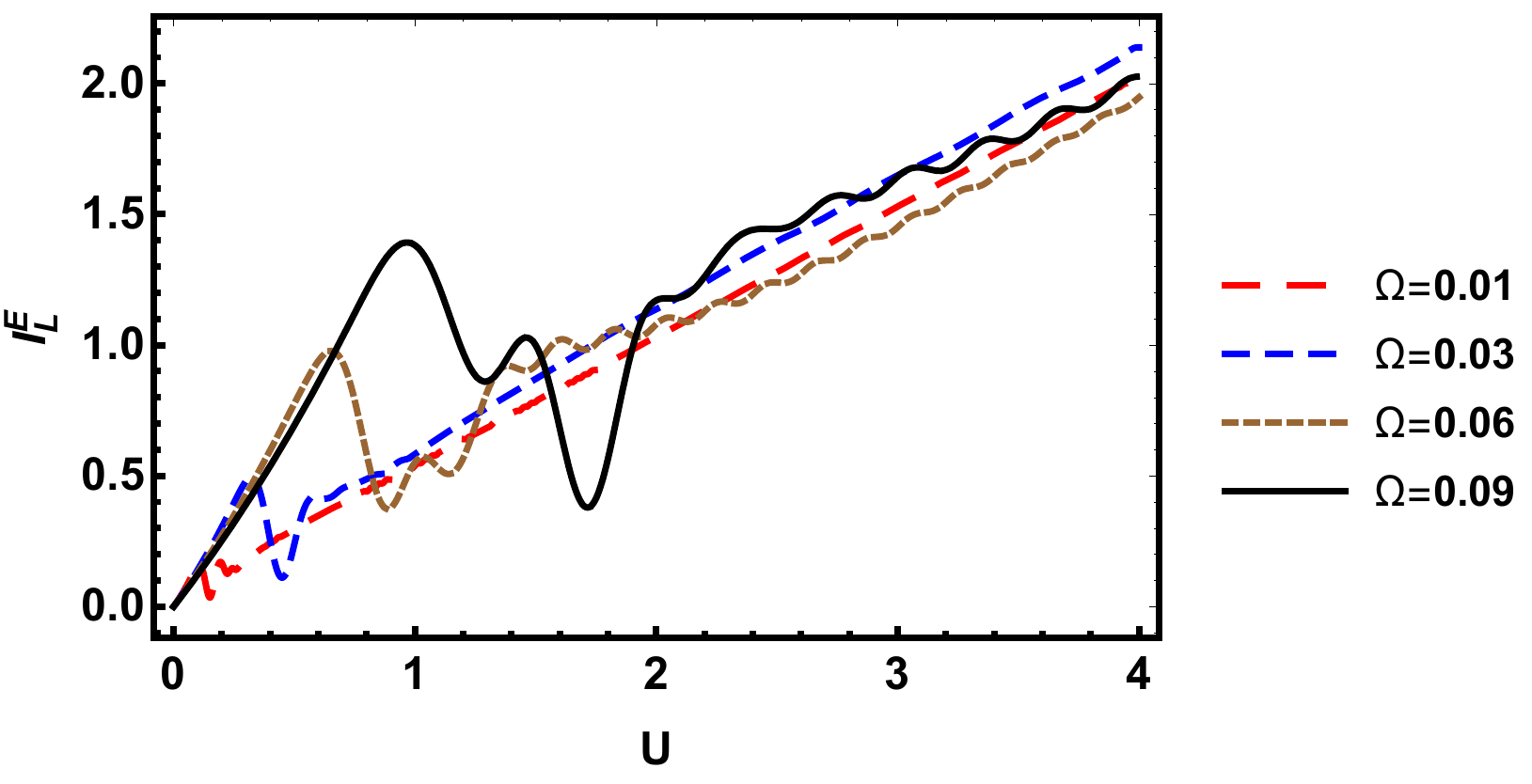}
	\caption{Up: Plot of left lead energy flux with U for three different values of the frequency $\Omega$ and $\mu=4,~t=5.$  The unit of $\Omega$ is same to the unit of $\Gamma$ and the unit of $t$ is same to that of $2\pi/\Omega.$}	
\end{figure}
In FIG. 4, the variation of the particle flux with $U$ is shown for very low frequencies of the drive. Although the variation is small for the smaller U values, the flux drops down to zero for larger $U.$ Contrary to the low frequency, the behavior changes completely for moderate and high frequency ranges, which depend very much on the time $t.$ 

Following the same trick, we write the power developed in the system by the source of time dependence(Eq. (\ref{37})) for common Fermi function i.e $f_{L}(\omega) =f_{R}(\omega)=f_{}(\omega) $ as follows
\begin{align}
&P_{d}(t)=-\frac{U\Omega}{2}{\alpha}_{0}^{Q}\sin[\Omega t]\nonumber\\&+\frac{U\Omega}{4}\sum_{l}\Big(\Delta\alpha_{l}^{Q}\sin[l \Omega t]
+\Delta\beta_{l}^{Q}\cos[l\Omega t]\Big),
\end{align}
where we have used
\begin{align}
\Delta\alpha_{l}^{Q}&=\alpha_{l+1}^{Q}-\alpha_{l-1}^{Q}\nonumber\\
\Delta\beta_{l}^{Q}&=\beta_{l+1}^{Q}-\beta_{l-1}^{Q}
\end{align}
and also using (\ref{alphaQ}) we have
\begin{align}
\alpha_{l\pm 1}^{Q}&=\sum_{m}\frac{2\Gamma }{\pi}\alpha_{m+l\pm 1,m}\nonumber\\
\beta_{l\pm 1}^{Q}&=\sum_{m}\frac{2\Gamma }{\pi}\alpha_{m+l\pm 1,m}.
\end{align}
In FIG. 5, we have shown plots for the power with $U.$ Here in the upper plot, we can see that in the very low frequency regime, the sign of the power is always -ve even if we increase time. However, for moderate frequencies, we have shown (in the down figure), that the power changes sign when time is varied. In the moderate frequency regime, thus we get change in the direction of the power as we change time. This is true for the time averaged power as well\cite{23}. In \cite{23}, the time average power is plotted for a high frequency $\Omega =2$ of the drive. Even for the averaged power, for $\Omega=0.03,$ the power as negative. As we increase frequency from $\Omega=0.03,$ the average power becomes $+ve.$ Similarly, in our case the direction of the power depends on the frequency of the drive and also on time $t.$ 
 \begin{figure}\label{fig7}	
 	\includegraphics[width=0.9 \linewidth]{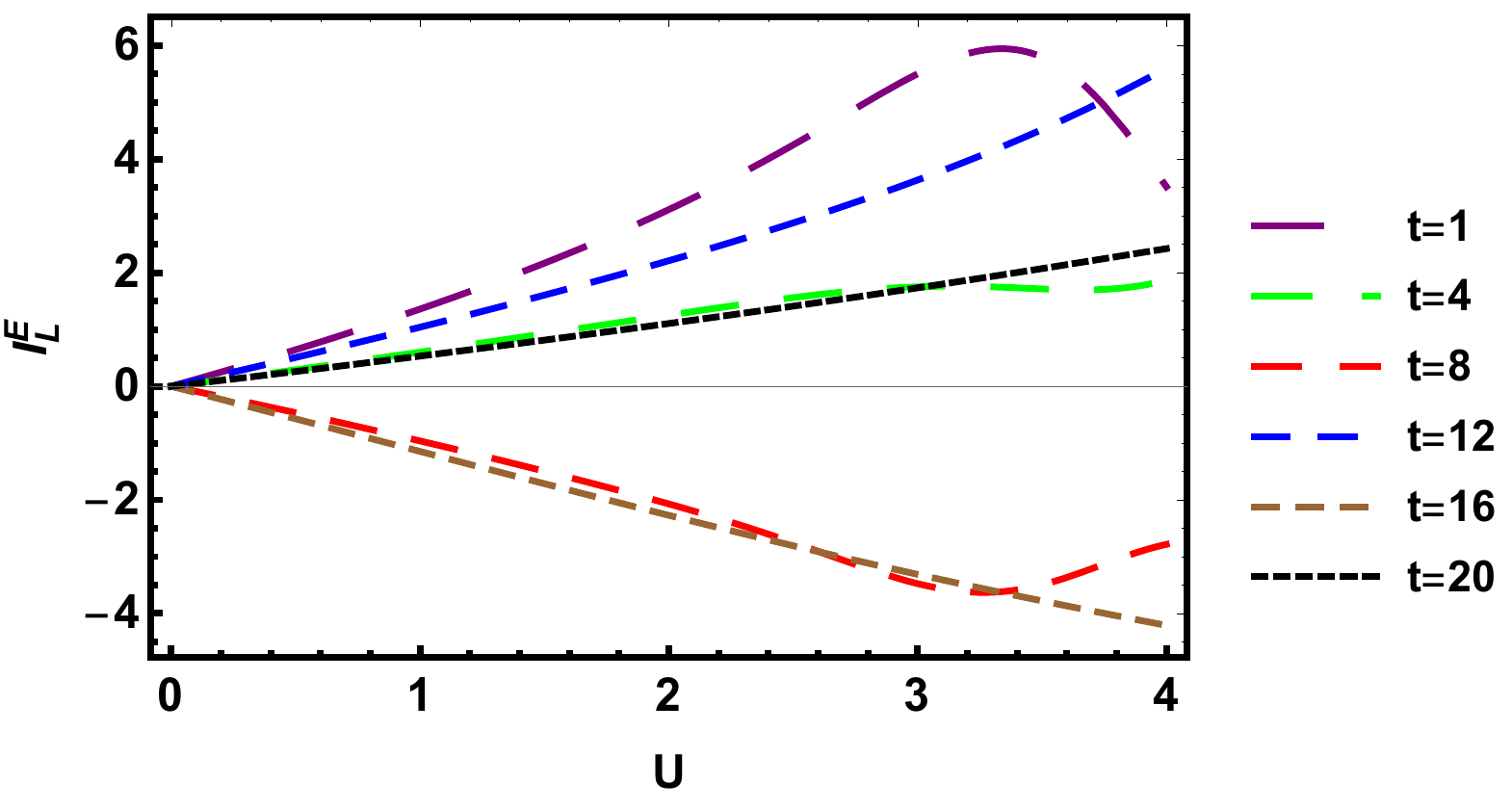}
 	\caption{Plot of left lead energy flux with $U$ for six different values of time $t.$ Here we have considered $\mu=4$ and $\Omega =0.3.$  The unit of $\Omega$ is same to the unit of $\Gamma$ and the unit of $t$ is same to that of $2\pi/\Omega.$}	
 \end{figure}
Finally, for the completeness we need a similar analysis for the left lead energy flux. From Eq. (\ref{ILE}), we write for $f_{L}(\omega) = f_{R}(\omega) =f(\omega),$
\begin{align}\label{IEt}
I_{L}^{E}(t)=I_{L}^{E,0}(t)+\sum_{l}\alpha_{l}^{E}\cos(l\Omega t)+\sum_{l}\beta_{l}^{E}\sin(l\Omega t),
\end{align}
 where
 \begin{align}\label{a1}
& I_{L}^{E,0}(t)=2\Gamma_{L}U\alpha_{0}^{Q} \cos(\Omega t)\nonumber\\&-\frac{2\Omega\Gamma_{L}}{\pi}\sum_{m} m J_{m}^{2}(x)\arctan(\frac{\mu-m\Omega}{\Gamma})
\end{align}
and 
 \begin{align}\label{a2}
 \alpha_{l}^{E}&=\Gamma_{L}U\Big(\alpha_{l-1}^{Q}+\alpha_{l+1}^{Q}\Big)-\Gamma_{L}\Omega\sum_{m} mJ_{l+m}(x)J_{m}(x)\nonumber\\&\times\Big(\frac{2}{\pi}\arctan(\frac{\mu-m\Omega}{\Gamma})-1\Big)\nonumber\\
 \beta_{l}^{E}&=-\Gamma_{L}U\Big(\beta_{l-1}^{Q}+\beta_{l+1}^{Q}\Big)-\frac{\Gamma_{L}\Omega}{\pi}\sum_{m} mJ_{l+m}(x)J_{m}(x)\nonumber\\&\log[\Gamma^{2}+(\mu-m\Omega)^{2}].
 \end{align}
 From Eqs. (\ref{a1}),(\ref{a2}) and (\ref{alphaQ}) one observes that the left lead energy flux is represented in terms of the coefficients used to discuss the left lead particle flux. The expression of the energy flux in Eq. (\ref{IEt}) provides the time dependent energy flux when the reservoirs are at common temperature(and also $T\rightarrow 0$) and common chemical potential$\mu$. In FIG. 6 we have presented the variation of the left lead energy flux with $U$ for a very low frequency of the time dependent field. In the low frequency($\Omega\sim 0.01$) range the energy flux is always positive and does not change sign when time is changed. On the other hand, when the frequency of the drive is increased($\Omega=0.3$), the energy flux changes sign from +ve to -ve and again -ve to +ve as we change time. For example in FIG.7, we have presented plot for which $t=1,4,8,12,16$ and the energy fluxes change sign. As long as we stay in the low frequency regime, our results will be unaltered as time changes. Another important aspect is that, we can show from Eqs. (\ref{IEt}), (\ref{a1}) and (\ref{a2}) that the right lead energy flux can be obtained by changing $\Gamma_{L}\rightarrow \Gamma_{R}.$ As evident from $(\ref{Gamma})$ that $\Gamma_{L}$ and $\Gamma_{R}$ are associated to the density of states of that leads, which lead would contribute much to the energy transport depends on how strong it is coupled to the dot.

\begin{figure}\label{fig8}	
	\includegraphics[width=0.9 \linewidth]{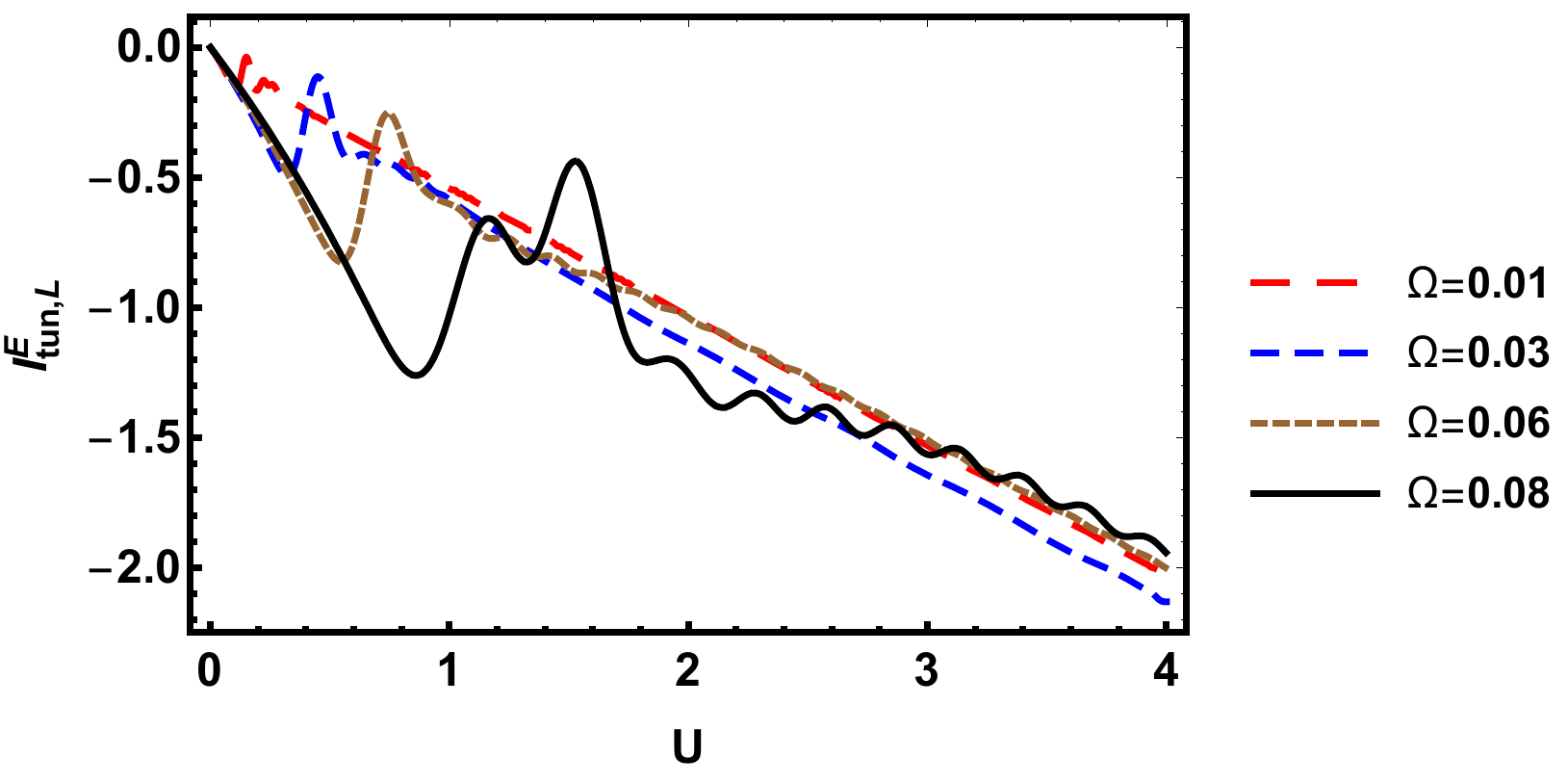}
	\caption{Plot of left lead tunneling flux with $U$ for different values of $\Omega.$ Here we have considered $\mu=4$ and $t=5.$  The unit of $\Omega$ is same to the unit of $\Gamma$ and the unit of $t$ is same to that of $2\pi/\Omega.$}	
\end{figure} 
  It is also important to comment here that the tunneling energy flux of the left lead under common chemical potential and common temperature of both reservoirs can be obtained as(see appendix B)
 \begin{align}\label{35}
 I^{E}_{{\rm tun},L}(t)=\epsilon^{}_{d}(t)I^{}_{L}(t)-I^{E}_{L}(t).
 \end{align}
This expression of the left lead tunneling energy flux shows that the knowledge of $I_{L}^{E}(t)$ and $I^{}_{L}(t)$ is enough to know the behavior of the tunneling flux. The left lead tunneling flux for the low frequency limit is plotted in FIG.8. As one can see it is -ve for this low frequency limit(and remains unaltered with change in time). With increasing frequency of the drive the sign of this flux  oscillates as time changes. Importantly, the above discussion includes both the regions of high frequency and low frequency of the driving. In \cite{lil,lil1}, the authors have represented $I^{E}_{{\rm tun},L}(t)/2$ as the energy reactance. This term can momentarily store energy in the time dependent transport phenomena. This term resembles to the electrical reactance, which also contributes in the time dependent transport. The energy reactance together  with the left lead energy current shows a positive heat current for all times for small or moderate values of $\Omega$ (when $\Omega<\Gamma$). Thus we define the effective energy current including the energy reactance as follows\cite{lil,lil1}
\begin{align}
I_{L}^{RE}(t)=I^{E}_{L}(t)+\frac{I^{E}_{{\rm tun},L}(t)}{2}.
\end{align}
In Fig. 9, we have presented the plot of the effective energy current $I_{L}^{RE}(t)$ with the driving parameter $U.$ Here for a frequency ($\Omega=0.07$), we always get a positive energy current for all given time.  In \cite{lil}, the authors have shown that the energy reactance provides a route to the universal response of a mesoscopic thermometer.
\begin{figure}\label{fig9}	
	\includegraphics[width=0.9 \linewidth]{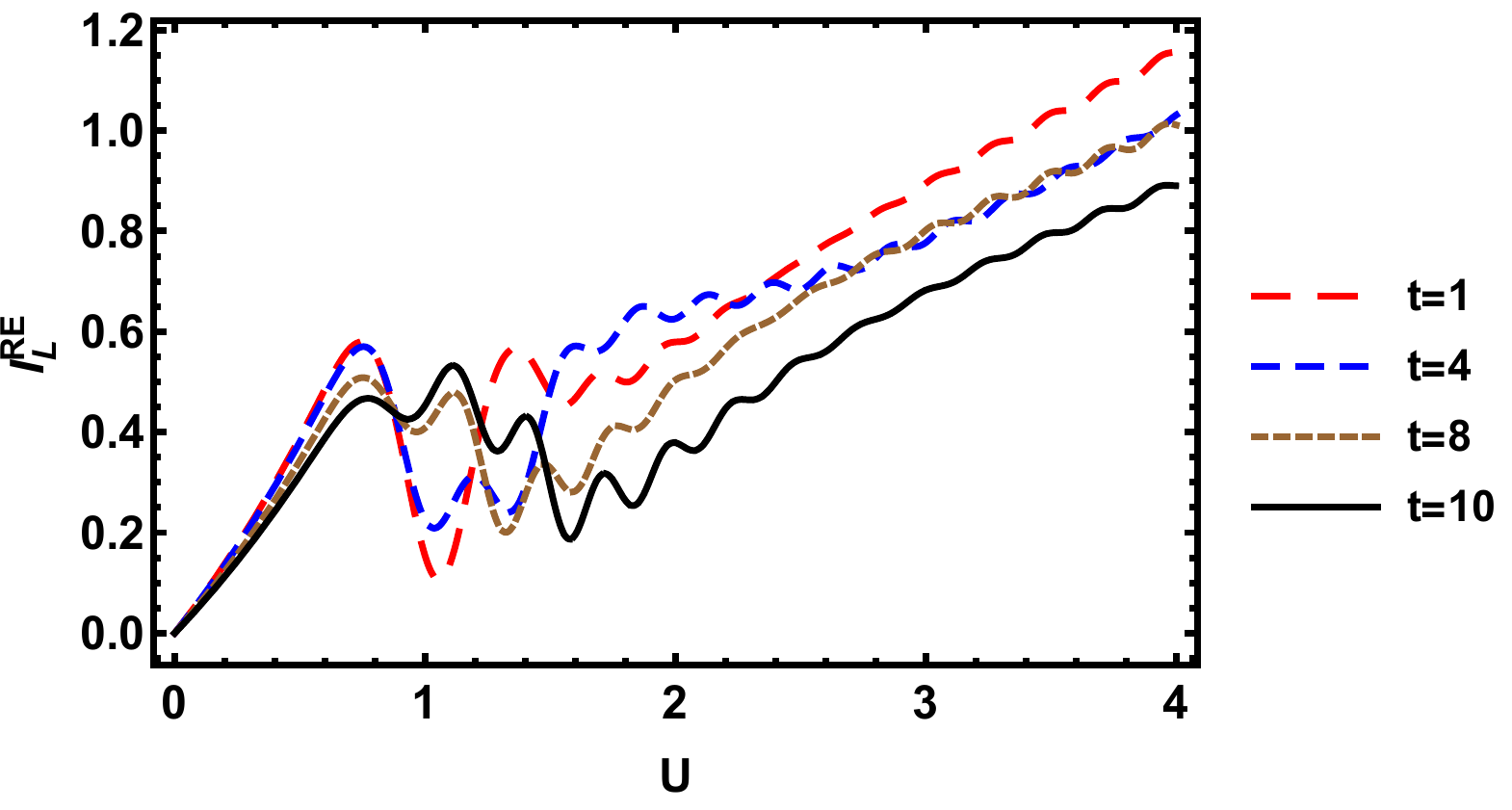}
	\caption{Plot of effective energy current with $U$ for different values of $t.$ Here we have considered $\mu=4$ and $\Omega=0.07.$  The unit of $\Omega$ is same to the unit of $\Gamma$ and the unit of $t$ is same to that of $2\pi/\Omega.$}	
\end{figure} 

Finally, let us discuss the feasibility of experiment for our model. In \cite{A}, the authors experimentally have demonstrated frequency-resolved impedance spectroscopy. A experimental set up is also proposed in ref. \cite{lil1}. Here the quantum dot is connected to a single reservoir at temperature $T.$ The dot is also capacitively coupled to a gate voltage, which provides periodic driving and is further attached to a floating contact. The floating contact is such that it can adjust time by time the chemical potential and temperature, associated to it, to avoid charge or heat accumulation in it. This unique proposal provides the experimental accessibility of our result as well.  In the low frequency limit of the driving, the heat currents could be measured by measuring the temperature of the floating contact with a thermometer. In our case, we need two reservoirs and they should kept at same temperature and chemical potential. When energy reactance term is added to the energy current it can be shown that the temperature of the floating contact does not vary with time and we get a steady temperature all the time.    
\section{Conclusion}
In this paper, we have considered an open quantum system containing a single level quantum dot , which is connected to two fermionic reservoirs via two metallic leads. The central part(the quantum dot) of the system is further perturbed by a time periodic electric field, which modifies the energy levels of the dot and makes the whole problem time dependent. In this time dependent system, the particle flux corresponding to a lead is computed considering the time derivative of the number operator of that particular lead. Furthermore, time derivatives of different parts of the Hamiltonian give energy fluxes related to that part of the system. Apart from these fluxes, the variation of the occupation number with time is presented. Similar treatment is done for the power delivered from the source of the time dependence towards the leads. One finds contribution from all harmonics of $\Omega,$ in the expressions of the dot occupation, power, particle and energy fluxes. In our problem we treat the two reservoirs to be at a common chemical potential $\mu$ and the temperature difference $T\rightarrow 0.$ In the low frequency regime of the applied field, we get a directed energy flow irrespective of the time. As we increase frequency, in the moderate frequency range, the direction of the energy flow can change sign depending on the time $t$. 

The novelty of this paper lies in the fact that it provides the time dependent fluxes in a simple form. As one can see in section IV, we have derived all time dependent quantities in a very simple form when the temperature and chemical potentials of the two reservoirs are the same and the common temperature $\rightarrow 0$. Within this scenario, it is possible to express the fluxes in terms of the different harmonics of the applied oscillatory field. The difference of the present paper with Reference \cite{23} is that, in \cite{23}, we had calculated the time averaged results of the fluxes(which are time independent and also contains integration over $\omega$ and time). From \cite{23}, although we had commented on the transmission, it was hard to comment on the behavior of the fluxes as those involve an integration over $\omega$ and the integrand involves the Greens functions as well as Fermi distribution functions. In the limit of common temperature and common chemical potentials of the two reservoirs and at $T\rightarrow 0,$ the Fermi functions become $\delta$ functions and this eventually help us to have simple forms of the fluxes. The presented and discussed results can be useful for planning and analyzing experiments in the fields of quantum thermodynamics and time-dependent quantum transport.

\section*{ACKNOWLEDGMENTS}

The author would like to acknowledge financial support from DST (project number SR/WOS-A/PM-52/2019). D.C would also like to thank Prof. Amnon Aharony and Prof. Ora Entin-Wohlman for suggesting this problem and collaborating during the early stages of this project. The initial work was done at Ben-Gurion university(BGU). Author is also indebted to BGU for local hospitality and its support towards research in basic sciences.

\appendix
\section{Necessary steps for the calculation of particle flux }
The particle and energy fluxes in the simple junction, is derived using the Keldysh Green's functions in the time domain. 
The particle flux into the left lead is obtained from the time derivative of the particle number of the left lead \cite{22,23}. Finally one can write the Particle current in terms of the Green's function of the dot as
\begin{align}\label{7}
I^{}_{L}(t)&=\langle \frac{d}{dt}\sum_{\bf k}c^{\dagger}_{\bf k}c^{}_{\bf k}\rangle\nonumber\\
&=\int dt^{}_{1}[\Sigma^{}_{L}(t,t^{}_{1})G^{}_{dd}(t^{}_{1},t)-G^{}_{dd}(t,t^{}_{1})\Sigma^{}_{L}(t^{}_{1},t)]^{<}_{}\ .
\end{align}
Here in Eq. (\ref{7}), $\Sigma^{}_{L}(t,t')$ is the self energy due to the tunnel coupling with the left lead,
\begin{align}\label{8}
\Sigma^{}_{L}(t,t')=\sum_{\bf k}|V^{}_{\bf k}|^{2}g^{}_{\bf k}(t,t')\ ,
\end{align}
with $g_{\bf k}(t,t')$ as the Green's function of the decoupled left lead and is defined as follows
\begin{align}\label{gk}
g^{r(a)}_{\bf k}(t-t')&=\mp i\Theta (\pm t\mp t')\langle\{ c^{}_{\bf k}(t),c^{\dagger}_{\bf k}(t')\}\rangle\nonumber\\&
=\mp i\Theta (\pm t \mp t')e^{-i\epsilon^{}_{k}(t-t')}\ ,
\end{align}
and
\begin{equation}
g^{<}_{\bf k}(t-t')=if(\epsilon^{}_{k})e^{-i\epsilon^{}_{k}(t-t')}\ ,\ \ \ f^{}_L(\epsilon^{}_{k})=\langle c^{\dagger}_{\bf k}c^{}_{\bf k}\rangle
\ ,
\label{gl}
\end{equation}
where we have indicated retarded (advanced) Green's  function by the superscript $r(a)$ and the corresponding change in the right hand side is the change of sign(corresponds to the upper (lower) sign on the right hand-side). In (\ref{gl}), 
\begin{equation}
f^{}_{L(R)}(\omega)=[e^{(\omega-\mu^{}_{L(R)})/(k^{}_{\rm B}T^{}_{L(R)})}+1]^{-1}
\label{Fermi}
\end{equation}
is the Fermi distribution function for the left and right lead. 
$G_{dd}(t,t')$ is the Green's function of the dot and is defined in Eq. (\ref{13}). The lesser Green's function of the product of two Green's function (as is there in Eq. (\ref{7})) is found by using the Langreth rule \cite{4} where for two Green's function $A$ and $B,$ one has $(AB)^{<}=A^{r}B^{<}+A^{<}B^{a},$ with $A^{r,a,<}$ are the retarded, advanced and lesser Green's functions respectively. The particle flux of the right lead can be obtained from Eqs. (\ref{7}) and (\ref{8}) by interchanging $L\Leftrightarrow R$ and ${\bf k}\Leftrightarrow {\bf p}$. 

Using the wide-band limit, in which the densities of states in the reservoirs are assumed to be independent of the energy \cite{1,2}, we have
\begin{eqnarray}
\Sigma^{r(a)}_{L}(t,t')&=\mp i\Gamma^{}_{L}\delta(t-t')\nonumber\\
\Sigma^<_{L}(t,t')&=2i\Gamma^{}_{L}\int\frac{d\omega}{2\pi}e^{-i\omega(t-t')}f^{}_{L}(\omega)\ .
\label{SIG}
\end{eqnarray}

Applying the Langreth rule, to Eq. (\ref{7}), and substituting Eqs. (\ref{SIG}), we find\cite{6}
\begin{align}
&I^{}_L(t)=2i\Gamma^{}_L\Big(\int\frac{d\omega}{2\pi}f^{}_L(\omega)\nonumber\\&\big[e^{-i\omega(t-t^{}_1)}G^a_{dd}(t^{}_1,t)-e^{-i\omega(t^{}_1-t)}G^r_{dd}(t,t^{}_1)\big]
-G^<_{dd}(t,t)\Big)\ .
\label{IL0}
\end{align}
The time-dependent particle current into the right lead is derived from Eq. (\ref{IL0}) by interchanging $L\Leftrightarrow R$ and ${\bf k}\Leftrightarrow {\bf p}$.

The Dyson's Eq. for the Green's function of the dot is,

	\begin{align}\label{13}
	G^{}_{dd}(t,t')=&g^{}_{d}(t,t')+\int dt^{}_{1}dt^{}_{2}g^{}_{d}(t,t^{}_{1})\Sigma^{}_{}(t^{}_{1},t^{}_{2})G^{}_{dd}(t^{}_{2},t') ,
	\end{align}
with
\begin{align}
\Sigma(t,t')=\Sigma^{}_{L}(t,t')+\Sigma^{}_{R}(t,t'),
\end{align}	
where $\Sigma_{L(R)}(t,t')$ is defined in (\ref{8}).
Here, 
\begin{align}\label{14}
g^{r(a)}_{d}(t,t')=\mp i\Theta (\pm t\mp t')\exp\Big [-i\int_{t'}^{t}dt^{}_{1}\epsilon^{}_{d}(t^{}_{1})\Big ]\ ,
\end{align}
which is the retarded(advanced) decoupled Green's function of the dot. Also,
$g^{<}_{d}(t,t')=0$, since the dot is assumed to be empty in the decoupled junction.
The retarded and advanced Green's function of the dot can be obtained from Eq. (\ref{13}) as
\begin{align}\label{15}
G^{r(a)}_{dd}(t,t')&=\mp i \Theta (\pm t\mp t')\nonumber\\&\times\exp\Big [-i\int_{t'}^{t}dt^{}_{1}\epsilon^{}_{d}(t^{}_{1})\mp \Gamma (t-t')\Big ]
\ .
\end{align}
Similarly the lesser Green's function of the dot is 
\begin{align}\label{16}
&G^{<}_{dd}(t,t)=i\int^{t} dt_{1}e^{-2\Gamma}(t-t_{1})\int^{t_{1}} dt_{2}\nonumber\\
&\times\Big(G_{dd}^{r}(t_{1},t_{2})\Sigma^{<}(t_{2},t_{1})-\Sigma^{<}(t_{1},t_{2})G_{dd}^{a}(t_{2},t_{1})\Big)
\end{align}
and thus
\begin{align}\label{lesGdd1}
G^{<}_{dd}(t,t')&=\int dt^{}_{1}\int dt^{}_{2}G_{dd}^{r}(t,t_{1})\Sigma^{<}(t_{1},t_{2})G^{a}_{dd}(t^{}_{2},t').
\end{align}
The electronic occupation on the dot can be obtained as
\begin{align}\label{17}
Q^{}_{d}(t)&=-iG^{<}_{d}(t,t)=
\int^{t} dt_{1}e^{-2\Gamma}(t-t_{1})\int^{t_{1}}\nonumber\\& dt_{2}\Big(G_{dd}^{r}(t_{1},t_{2})\Sigma^{<}(t_{2},t_{1})-\Sigma^{<}(t_{1},t_{2})G_{dd}^{a}(t_{2},t_{1})\Big).
\end{align}
Differentiating Eq. (\ref{17}) with respect to $t,$ we get
\begin{align}\label{18}
-\frac{dQ_{d}(t)}{dt}&=2\Gamma Q_{d}(t)-\int^{t}dt_{1}\Big(G_{dd}^{r}(t_{1},t_{2})\Sigma^{<}(t_{2},t_{1})\nonumber\\&-\Sigma^{<}(t_{1},t_{2})G_{dd}^{a}(t_{2},t_{1})\Big),
\end{align}
Using Eq. (\ref{18}) in (\ref{19}) we get Eq. (\ref{20}).\\

\section{Derivation of Equation (\ref{35})}\label{B}
The temporal variation of the tunneling part of the Hamiltonian provides 
tunneling energy flux as follows
\begin{eqnarray}
&I^{E}_{{\rm tun}, L}(t)
=\epsilon^{}_{d}(t)I^{}_{L}(t)-I^{E}_{L}(t)\nonumber\\
&+\sum_{{\bf k},{\bf p}}[V^{\ast}_{\bf k}V^{}_{\bf p}G^{<}_{{\bf k}{\bf p}}(t,t)-V^{}_{\bf k}V^{\ast}_{\bf p}G^{<}_{{\bf p}{\bf k}}(t,t)]\ ,
\label{B1}
\end{eqnarray}
The Dyson Eqs for $G_{{\bf kp}}$ and $G_{{\bf pk}}$ are obtained as
\begin{widetext}
\begin{eqnarray}\label{B2}
&G_{{\bf kp}}(t,t')=V_{{\bf k}}V_{{\bf p}}^{*}\int dt_{1}\int dt_{2}g_{{\bf k}}(t,t_{1})G_{dd}(t_{1},t_{2})g_{{\bf p}}(t_{2},t')\nonumber\\
&G_{{\bf pk}}(t,t')=V_{{\bf p}}V_{{\bf k}}^{*}\int dt_{1}\int dt_{2}g_{{\bf p}}(t,t_{1})G_{dd}(t_{1},t_{2})g_{{\bf k}}(t_{2},t')\ .
\end{eqnarray}
\end{widetext}

Using (\ref{B1}),(\ref{B2}) and (\ref{ITLE}), we finally can write the tunneling energy in the following form
\begin{widetext}
\begin{eqnarray}\label{B3}
I^{E}_{{\rm tun},L}(t)
&=&\epsilon^{}_{d}(t)I^{}_{L}(t)-I^{E}_{L}(t)\nonumber\\&+&2\Gamma_{L}\Gamma_{R}\int \frac{d\omega}{2\pi}\Big(f_{R}(\omega)-f_{L}(\omega)\Big)\int dt_{1}\Big[e^{-i\omega(t_{1}-t)}G_{dd}^{r}(t,t_{1})+e^{-i\omega(t-t_{1})}G_{dd}^{a}(t_{1},t)\Big]\ .
\end{eqnarray}
\end{widetext}
For common Chemical potential of the two reservoirs, from Eq. (\ref{B3}) we get Eq.  (\ref{35}) of the main text.
\end{document}